\documentclass[twocolumn]{svjour3}          
\smartqed  
\usepackage{graphicx}
\usepackage{lipsum}
\usepackage{url}
\journalname{KI - K{\"{u}}nstliche Intelligenz}

\usepackage[ngerman]{babel}
\usepackage[ansinew]{inputenc}
\usepackage{umlaute}
\usepackage{wrapfig}

\def\makeheadbox{\relax}

\begin{document}
{\sloppy

\title{Quantum Technologies and AI\\
{\large Interview with Tommaso Calarco} 
}

\titlerunning{Interview with Tommaso Calarco}       

\author{ Matthias Klusch \and J\"org L\"assig \and 
         Frank K. Wilhelm}

\authorrunning{Klusch, L\"assig \& Wilhelm} 

\institute{Matthias Klusch \at
              German Research Center for Artificial Intelligence GmbH (DFKI), Saarbr\"ucken, Germany\\
              \email{matthias.klusch@dfki.de}           
           \and
           J\"org L\"assig \at
              University of Applied Sciences Zittau/G\"orlitz, Germany\\
              \email{jlaessig@hszg.de}
           \and
           Frank K. Wilhelm \at
           Forschungszentrum Juelich GmbH (FZJ), J\"ulich, Germany
           \email{f.wilhelm-mauch@fz-juelich.de}
}

\date{ }

\maketitle


\begin{center}
\includegraphics[width=0.6\columnwidth]{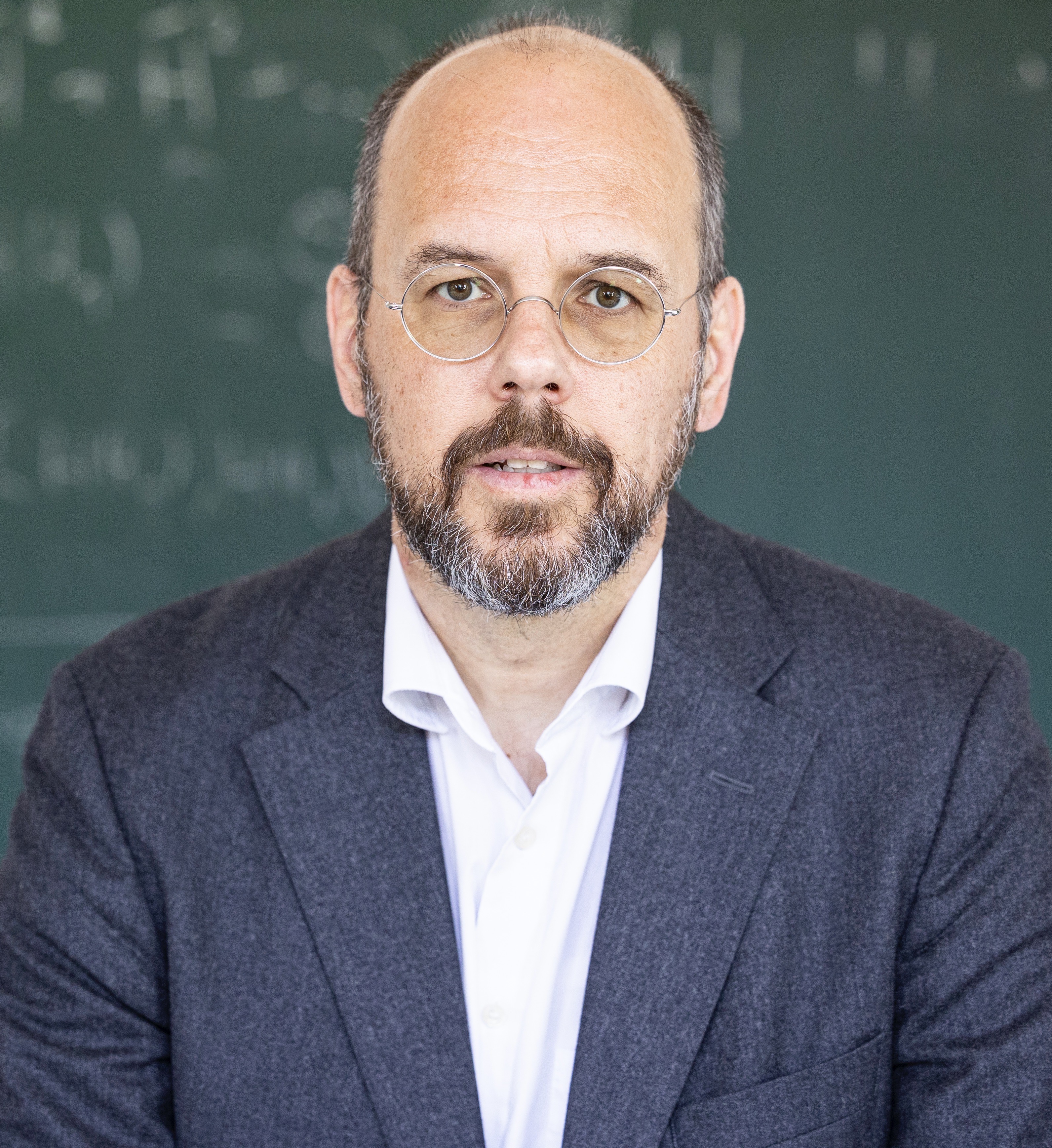}
\end{center}
 
\noindent 
Prof. Dr. Tommaso Calarco has been a full professor at the University of 
Cologne since 2018 and at the University of Bologna since 2023. He received 
his PhD at the University of Ferrara and then to worked as a postdoc in the 
group of Peter Zoller at the University of Innsbruck. He was appointed as a 
Senior Researcher at the BEC (Bose-Einstein Condensation) Centre in Trento in 
2004 and as a Professor for Physics at the University of Ulm in 2007, where 
he then became Director of the Institute for Complex Quantum Systems and of 
the Centre for Integrated Quantum Science and Technology. He has authored in 
2016 the Quantum Manifesto, which initiated the European 
Commission\textquotesingle s Quantum Flagship initiative, and is currently 
the chairman of one of the Flagship\textquotesingle s governing bodies: The 
Quantum Community Network (QCN). In 2020, together with the QCN, he has 
launched an initiative towards the creation of a consortium of European 
quantum industries, which has been legally established in 2021 under 
the name of European Quantum Industry Consortium (QuIC).
\\[2ex]

\section{Introduction} 

\textit{KI: Tommaso, thank you very much for being available for this 
interview! Let us first briefly introduce you to the readers of this AI 
journal: How did you enter the field of quantum technologies? 
What were and are your main personal motivations to work in the field?
}\\

\noindent
Rather than me entering the field of quantum technologies, 
the field entered into my life when it was not even called that way 
but foundations of quantum mechanics. I first became interested in this field 
when doing my diploma thesis, and then for several years I kept a position as 
a elementary school teacher and then a high school teacher as a sort of 
fallback option. Because at that time, I was pretty sure that I would never 
have a career working in such a fascinating but utterly useless field as 
quantum mechanics; in fact, everybody was telling me that if I will do 
then I would ruin my career. But soon after that quantum technology started 
entering the world and our lives with the discovery of the first quantum algorithms, 
the Shor algorithm and the Grover algorithm, together with the first gate-based 
implementations of quantum computers. From that point on, it became clear to me that 
quantum mechanics was not only fun but also useful, such that perhaps a career could 
eventually come out of it, which then indeed happened to me.\\

\noindent
\textit{KI: How would you define useful in this context?}\\

\noindent
I would say that useful in the trivial sense is something where I can build a 
device which can be used for some purpose but rather outside the science it 
is originated in. For instance, my diploma work was about the Bell 
inequalities, something of great use to reflect in depth on whether the world 
exists locally or not. Which is as philosophical as it gets for any question 
and brought Alan Aspect, Anton Zeilinger and John Clauser the Nobel Prize in physics 
(in 2022 for showing that the quantum world is not local in real through his 
experiments in the early 1980s, which validated the principle of quantum entanglement 
by violation of Bell's inequalities).
But the usefulness of that was sort of, say, purely scientific in terms of advancing 
knowledge only without any broad application of it.\\

\noindent
\textit{KI: What are your current main activities in quantum 
technologies, both scientifically and as a community leader?}\\

\noindent
Well, I have started several years ago with my research to leverage ideas from 
machine learning and AI, especially reinforcement learning to improve the 
performance of quantum devices. I have founded a department of the 
Research Center Julich that focuses on this subject, called automated 
optimization and quantum optimal control. On a broader level, I am trying to connect the dots and keeping 
together the community, for example, by chairing the quantum community 
network, which is one of the three governance bodies of the European quantum 
flagship. And at the same time, I am doing my best to connect what the 
community is thinking, conceiving and seeing as the future vision towards 
decision makers, especially at the European, but also at the national level
with different kinds of strategies, documents, events and initiatives. 

\section{Quantum Technologies} 

\textit{KI: You are well-known for your work in quantum technologies, 
which indeed made quite some headlines particularly over the last years. 
But are quantum technologies really a new thing - or what is now different 
compared to a few decades ago?}\\ 

\noindent
Let me give some sort of quantum answer, yes and no. There is nothing conceptually new, 
because all of these concepts used in quantum technology, such as superposition and 
entanglement, are actually something which was already present and very much at the 
core of the theory of quantum mechanics since its inception. But then what is new? 
We only started to realize in the 1980's at the earliest that quantum mechanical 
systems can be used for the purpose of processing information. 
Richard Feynman and David Deutsch started asking whether we can use quantum 
mechanical objects to process information. This was the birth of quantum information, 
which is not new physics, maybe not even really a new theory. 
But it is indeed very much a new way to look at those known quantum phenomena 
and combining them, which ultimately leads to interdisciplinary links with 
information theory and computer science. 
In addition, we can do way more with quantum computing resources available to us
nowadays than the founding fathers of quantum mechanics thought would ever be 
possible. For example, Erwin Schr\"odinger said that it is ridiculous to even 
imagine to experiment with a single atom in thought experiments, though we often 
do that, as this always entails ridiculous consequences; it would be like the
idea of raising a pterodactyl in a zoo. Since then, huge progress was made in 
the refinement of capabilities to manipulate individual quantum objects, which 
allows you to also use them for processing information. This is what really led 
to this development of quantum technologies.\\

\noindent
\textit{KI: In the early 2000s, quantum computing was one prominent research 
topic to work on even in funded projects. Is the situation today really so much
different from back then?}\\

\noindent
Well, again, yes and no. It is different in the sense that 15 or even 30 years 
ago, we thought it was a dream. When I started teaching my course on quantum 
information more than 15 years ago in Ulm, I said to students: nice ideas but they are
more or less science fiction. But today, from the physics point of view, the 
situation became substantially different in that we now have quantum computers with 
dozens of quantum bits such that it might become really possible to do these 
things from the technology or applications point of view. Though we do not yet 
have a quantum computer which can really do something qualitatively different in 
the sense that no classical supercomputer can perform, I think, we are well on 
our way there. In this respect, a cautious but strong optimism is justified.\\

\noindent
\textit{KI: From your perspective, how are Europe and Germany positioned in this 
field today, technology-wise and regarding research funding and investments? 
Does Europe lag behind in this respect compared to the current major players 
in the field, USA and China?}\\

\noindent
Short answer: In terms of private investment yes, but public funding no, 
in the sense that Europe is the highest public net investor with around 9 billion 
Euros in quantum. The US is not investing so much, at least not in the open.
Private capital is just the reverse: Way more venture capital and private investment 
are in the US. I think, that is also a major risk we have at the moment, 
in terms of sort of limited ability from our startups to really scale up.
In addition, there is the risk that foreign capital is acquiring our startups 
and essentially doing what happened with the Internet having been invented in Geneva 
by physicists but the whole money being elsewhere. 
I did ask the head of Google quantum AI lab what he thinks we can do to stimulate a 
comparable level of investment in Europe as we see in the US from the private 
companies? The straight answer was that we cannot because we do not sit on a pile 
of cash.

Now, consider, for example, the Samsung Galaxy Quantum smartphone which 
has the simplest possible quantum device inside, that is a quantum random number 
generator, designed to secure sensitive transactions. The generator gets a qubit in 
a superposition zero plus one, and measures it, and then you get a result with 50\%  
probability. 
This is the first consumer electronic device that contains quantum technologies of 
the second quantum revolution, though, without a quantum computer inside and 
no quantum communication. This quantum device has been developed by the Swiss company 
ID Quantique (IDQ) in Europe but South Korea Telecom now owns the majority of shares 
for IDQ, the company producing it. So, I think, we should avoid that foreign capital 
can come in and acquire our companies, thereby depriving our ecosystem of the not 
only technological, but also value chain creation sovereignty, which is necessary to 
fuel the whole system and generate more venture capital and be able to invert the 
tendency. That is one main challenge that we have now.\\

\section{Quantum AI} 

\textit{KI: Let us talk about quantum AI which combines quantum technologies with AI.
From your perspective, what are the general relations between both disciplines,
and how would you assess the current status of quantum AI, regarding quantum 
computing for AI, and AI for quantum computing?} \\

\noindent
Exactly, quantum AI goes into two directions: One refers to the 
the use of classical AI for different tasks which help in building and 
operating quantum computers. For instance, one could leverage reinforcement 
learning for optimizing the control, performance, calibration and 
characterization of quantum devices in order to suppress errors and drive our
systems in the best possible, coherent and quantum way. 

There are other applications of AI in the field of quantum 
technologies which also go on a higher level in the stack in terms of, 
for instance, synthesizing and compiling algorithms for quantum computers. 
For a certain algorithm one could synthesize a gate sequence which sort of 
breaks it down into assembler language, in such a way that you get 
it in the most efficient possible way. One thing that we might envision is 
using AI in order to develop a search for new algorithms, potentially because 
we do not have a systematic theory of how you can synthesize or find 
algorithms. The limited number of quantum algorithms that we have are more or 
less anecdotal, and even the people who proposed them cannot explain in 
a systematic way how they came to the idea. So we do not have a general 
understanding of that, such as we have in classical programming. 

And there are more fields in which you can use AI in the context of quantum 
computing such as for simulation, classification, and investigation of 
many-body quantum states, which is relevant for quantum simulation. For 
example, in my institute, we have a young investigator group which is 
oriented towards using so called neural quantum states. In other words, 
AI-inspired classification to calculate computer dynamics of many-body 
states. There is really a host of different applications of AI within 
quantum. And, of course, one idea that we are also considering is, would it 
make sense to have a large language model (LLM) which is dedicated to quantum? 
The European Commission is considering those options, and research is sort of 
very diverse in this one direction of quantum AI: AI for quantum computing. 

The other direction of QAI refers to the use of quantum computing for AI. 
Now, is there really such a link? Well, from my perspective, there is one  
which is conceptually very clear and refers to a computational speedup of 
quantum search algorithms of which Grover's algorithm for search in a 
structured database is a popular example. 
Another link is adiabatic quantum computing, or quantum annealing for 
complex optimization problems in AI. There is a hope that quantum annealers 
can help now to reach production capability. This would require the capacity 
of writing and reading in quantum memory, like having a huge number of qubits 
which can be kept coherent with error correction in a recursive way to achieve 
fault tolerance. We need coherent qubit operations, even in the presence of errors
over a very large number of qubits combined and error corrected, with a lower error 
rate, and that needs to be scaled in a recursive way. So, the optimization aspect 
is one clear direction and bridge between quantum and AI.\\

\noindent
\textit{KI: In the past couple of years, a tremendous progress has been made
in showing quantum utility of direct quantum or hybrid quantum-classical 
algorithms for solving certain computationally hard problems over their 
classical counterparts in AI. In this regard, quantum AI is not an 
academic moon shot anymore. But, from your perspective, what level of 
maturity has QAI reached? Do you envision sufficient commercial 
potential of what kind of real-world QAI applications that, in turn, may 
drive research and software development in this field even further?}\\

\noindent
I agree with you that there has been a lot of progress on the 
software side and on the theory side in this field. Whether and when 
this can be commercialized and used in practice at large  
is literally the multi-billion dollar question, in the sense that it 
completely hinges on the progress of quantum hardware development.
The quantum computing resources needed to implement those QAI algorithms 
are really beyond NISQ (noisy intermediate scale quantum computers),
and NISQ is the stage in which we are mostly finding ourselves with a few 
dozens of physical qubits today. 

Some people argue that the recent result by Harvard with 48 logical qubits 
\cite{H1} takes us one step beyond NISQ towards the real age of quantum 
computing. But we need full scalability in the hardware for those ideas to 
become practically relevant, in the sense that you can not only think that 
there is an advantage, but you can actually deploy it. 
In a sense, it is reminiscent of classical AI in that a lot of ideas for, 
for example, deep learning were already present before the required sheer 
computational power and data became available. 

So, there are a lot of amazing developments nowadays, but conceptually, 
a lot of what we see today is building on a foundation which was already 
laid some time ago. But, you know, then the first and also the second AI 
winter came just because the hardware could not keep up with these wonderful 
ideas; so, in this sense, I do see an analogy here. 
But yes, the potential is there, and, yes, the conceptual maturity of quantum AI 
is there. But we need, you know, those engineers in quantum computing and AI 
which will make these things real. \\

\noindent
\textit{KI: One might argue for not lagging behind in or even giving up on 
research and development of quantum AI algorithms just because quantum 
computers are not yet ready enough for their deployment at large scale. 
What is your opinion in this regard?}\\

\noindent
Yes, I think, that is a very, very valid and important statement!
As a matter of fact, it has always been the case that the amazing 
developments on the algorithm side have been really a strong pull 
for the hardware to come along.
Without even those few early quantum algorithms like Shor's factoring 
algorithm, we would not be where we are now with quantum computing.
Because what it promises or threatens is to break conventional encryption schemes 
was certainly a big driver for funding research and pushing for the development of 
the hardware. So, absolutely it makes sense and we should develop further 
those ideas of quantum AI even though the actual deployment on  
quantum computers at large scale may be a few years down the line. \\

\noindent
\textit{KI: Apart from showing the feasibility of quantum AI algorithms for solving 
selected hard AI problems, investigating their potential of quantum utility, 
rather than quantum advantage, is of particular interest. This activity is 
complemented with the development of appropriate (non-NISQ) quantum computing 
devices which would become available when, in 10 or 20 years from now? }\\

\noindent
I would indeed say between 10 and 20 years from now, as that is also the 
expectation in the quantum community today. And I would like to come back 
to one expression that you used, which is very, very important in this 
context, which is the concept of quantum utility. Very correctly, you 
contrasted it with quantum advantage, because quantum advantage has long 
been, with good reasons, of course, the holy grail of the scientific 
community in terms of asymptotic advantage. 
Like, I build a quantum machine for something a classical supercomputer 
will never be able to reproduce. That is, of course, fantastic, fascinating,
a wonderful dream, and we know it is nowadays realizable for some very specific,
not practically relevant problems. But industry might be very well interested in 
something which, from the theoretical computer science perspective might be 
completely boring, like 'only' a factor of ten improvement in your 
computing power already bringing a lot of value, then a hundred fold improvement 
is enormous. So this aspect of quantum utility, which is the practical, 
the pragmatic aspect, is really what we should be focusing on now, 
because that should be not like the holy grail, but really the carrot hanging 
from the stick, in a sense.\\

\noindent\textit{KI: 
Regarding the various subfields of AI the field of quantum AI is not 
just restricted to quantum machine learning as it is still widely assumed 
by many. But: How to address what kind of potential problems the 
strong interdisciplinarity of QAI might impose on scientists from the 
originally separate fields of quantum physics and computer science?} \\

\noindent
In this regard, at the moment, there is the, may I say, kind of a little bit 
old fashioned physicists way of looking at things: There are those engineers, 
those computer scientists to be evangelized about the holy world of the 
quantum, and then they will be enlightened. No, what we need is 
really a two-way education. This is in my little corner what I am trying to 
achieve for myself and for my institute. We need a lot more interdisciplinarity. 
We need, as a scientist, a lot more conferences, projects, more activities 
on that subject QAI. As a matter of fact, I am aware that the European Commission is  
discussing about starting a seed activity on quantum and AI in both directions, 
to start laying the ground for this. \\

\noindent
\textit{KI: So, it appears that interested scientists from related 
disciplines such as computer science, AI and quantum physics still would 
need to somehow develop a common culture, mentality and standard of working 
together in quantum AI. But what are the currently most promising directions 
and challenges of research and development in QAI?
}\\

\noindent
Well, first of all, certainly developing a common language, like a glossary 
of terms which are known to both computer scientists and quantum physicists. 
Second, developing a reciprocal understanding of, really, the potential. 
Third, I would say, identifying cases of application, 
developing a systematic and deep understanding of all the possible 
implications of the acceleration, which can come from the few problems that we 
know how to solve with algorithms on a quantum computer. Where is it that 
they can provide utility, not necessarily advantage, as you said before, a 
systematic investigation of that. I am not aware that this exists, though 
there are a lot of points of evidence. 

The other challenge is really a thorough understanding of the resources 
associated, not to fall back to the often used argument $"$Oh no, this is not 
realistic, so let us forget it$"$. Rather we need to be able to prioritize, 
to understand what are the low hanging fruits in this respect, what are the 
AI applications which cost the least quantum resources per utility. 
That is, we need a systematic investigation accompanied with the hardware challenge 
of making sufficiently stable quantum memory and so on.\\

\noindent\textit{KI: And finally, is there anything on the potential 
benefits and risks associated with development of quantum technologies and AI 
that we do not have on our radar both as a community and we as interviewers 
that you would like to emphasize?}\\

\noindent
Well, yes, there are two risks: Again, one practical risk is a foreign direct 
investment and acquisition of quantum AI technologies taking over our own 
quantum technology ecosystem, and the other one is the dual use of it. 
We know that some countries such as Spain, France and UK have started
listings of export control regulations for quantum computers with certain 
number of qubits and error rates and so on, to counteract the fear that 
the capabilities of quantum computers can be potentially abused by 
non-likeminded countries.\\

\noindent
\textit{KI: What advice would you give young researchers who want to enter 
the field of quantum AI: How to start, where to start?}\\

\noindent
Oh, well, if they want to enter, there is just a plenty of opportunities. 
First, they should know that they are very much welcomed in the sense that there is 
a strong need for such people, not just but in particular in the context of funded 
quantum AI research projects. They should just look, for example, 
for opportunities on the community website of the European Quantum Technology 
flagship \cite{H2}. They should not be intimidated
by the purported difficulty of quantum computing. Actually, within a few weeks, 
with reading, for example, just a few chapters of the book of Nielsen and Chuang 
\cite{H3}, you can get to know enough about quantum mechanics to start working with 
the basic building blocks of quantum computing here. So they should just engage and 
not be afraid!}\\

\noindent
\textit{KI: Tommaso, thank you very much for your time and this very 
enlightening interview!}



\begin{thebibliography}{}
%
\bibitem{H1}
Bluvstein, D. et al. (2024):
Logical quantum processor based on reconfigurable atom arrays. 
{\em Nature}, 626(7997). 
Also see: 
\url{phys.org/news/2023-12-logical-qubits-quantum-errors.html}

\bibitem{H2}
European Quantum Technologies Flagship: \\
\url{https://qt.eu/}

\bibitem{H3}
Nielsen, M.A. \& Chuang, I.L. (2010):
Quantum computation and quantum information. 
Cambridge university press.

\end{thebibliography}
\end{document}